# A Novel Countermeasure Technique to Protect WSN against Denial-of-Sleep Attacks Using Firefly and Hopfield Neural Networks (HNN) Algorithms

Reza Fotohi[1]  . Somayyeh Firoozi Bari[2]

**Abstract**

Wireless Sensor Networks (WSN) contain numerous nodes that their main goals are to monitor and control environments. Also, sensor nodes distribute based on network usage. One of the most significant issues in this type of network is the energy consumption of sensor nodes. In fixed-sink networks, nodes which are near the sink act as an interface to transfer data of other nodes to sink. This causes the energy consumption of sensors reduces rapidly. Therefore, the lifetime of the network declines. Sensor nodes owing to their weaknesses are susceptible to several threats, one of which is Denial of Sleep Attack (DoSA) threating WSN. Hence, the DoSA attack refers to the energy loss in these nodes by maintaining the nodes from entering energy-saving and sleep mode. In this paper, a hybrid approach is proposed based on mobile sink, firefly algorithm based on leach, and Hopefield Neural Network (WSN-FAHN). Thus, mobile sink is applied to both improve energy consumption and increase network lifetime. Firefly algorithm is proposed to cluster nodes and authenticate in two levels to prevent from DoSA. In addition, Hopefield Neural Network detects the direction route of the sink movement to send data of CH. Furthermore, here WSN-FAHN technique is assessed through wide simulations performed in the NS-2 environment. The WSN-FAHN procedure superiority is demonstrated by simulation outcomes in comparison with contemporary schemes based on performance metrics like Packet Delivery Ratio (PDR), average throughput, detection ratio, and network lifetime while decreasing the average residual energy.

**Keywords** Wireless Sensor networks (WSNs) . Denial of Sleep Attack . Firefly algorithm . Mobile Sink . Hopefield Neural Network

✉ Reza Fotohi
 R_fotohi@sbu.ac.ir; Fotohi.reza@gmail.com

✉ Somayyeh Firoozi Bari
 S.firoozi84@gmail.com

[1] Faculty of Computer Science and Engineering, Shahid Beheshti University, Tehran, Iran
[2] Department of Computer Engineering, Shabestar Branch, Islamic Azad University, Shabestar, Iran

# 1 Introduction

**W**SN includes a set of self-guiding sensors monitoring the circumstances like sound, pressure, vibration, and temperature. The energy for the WSN's sensor nodes is supplied by batteries. However, energy loss is one of the main problems in WSN [1] that is resultant from overhearing, collisions, listening, idle, control packet overhead. The energy loss in the collision loss is introduced by the data packets' collision in the wireless medium. The radios maintenance presents the energy loss in the overhearing loss in the receiving mode over data packet transmission. A radio of the node in just monitoring the channel creates the idle listening loss. As the control packets should be received by all the nodes in the transmission range, it leads to introducing the control packet overhead. In general, the WSN is susceptible to two kinds of attacks as invasive and non-invasive attacks. The power, timing of the channel, and the frequency are affected by the non-invasive, however, information transmission, service availability, and routing process are influenced by the invasive attacks. Among the WSN attacks, the service or the system is made inaccessible by DoSA [2]. The DoSA is one of the WSNs' power exhausting attacks as a special Denial-of-Service (DoS) trying to maintain the sensor nodes awake to use further energy of the controlled power supply. In addition, the energy consumption of the sensors increases in DoSA attack by preventing the sensor from sleeping [3, 4]. The attacker node can transfer dummy data packets to authorized nodes, which causes an unnecessary transmission as well as increasing the energy consumption. It also processes the data obtained from the attacker nodes if the receiver sensor cannot detect the source when receiving data packets. Therefore, the energy consumption extremely increases in processing procedures and the sensors run out of energy.

In recent years, researchers have studied how to use mobile sink in networks. The use of mobile sink in networks causes the distance between nodes from the sink reduces. The reduction of the distance makes the energy consumption decreases. Therefore, the lifetime of the network increases. Applying hierarchical algorithm on networks is a proper way to improve energy consumption (decrease energy consumption) [5-7]. Thus, in this paper, we examine the movement of the sink in hierarchical networks which consist of some clusters. Each cluster consists of some typical nodes that receive needed data and forward them to CH. Proposed method has four phases. Clusters are made with fire-leach algorithm based on leach protocol in first phase. In second phase, data of each cluster is sent to determined CH after authentication at cluster level. Then, optimal points are considered by Hopefield Neural Network (HNN) in third phase. In addition, the mobile sink is implemented based on the hierarchical algorithm for routing direction in WSN. The direction of the sink movement is determined according to $T_{dr}$ deadline at each point with the radius of transmission of CH. $T_{dr}$ time has a significant (noticeable) effect on the HNN design. Also, in fourth phase, the authentication of the sent data is done by CH to the mobile sink. In fixed sink sensor networks, the nodes which are near the sink act as an interface to send data of other nodes to the sink. This causes energy consumption of nodes reduces. The energy consumption decreases greatly by applying mobile sink in the network because the sink is able to move between sensor nodes. The DoSA scenario is shown in Figure 1.

**Fig. 1** WSNs with DoSA attack.

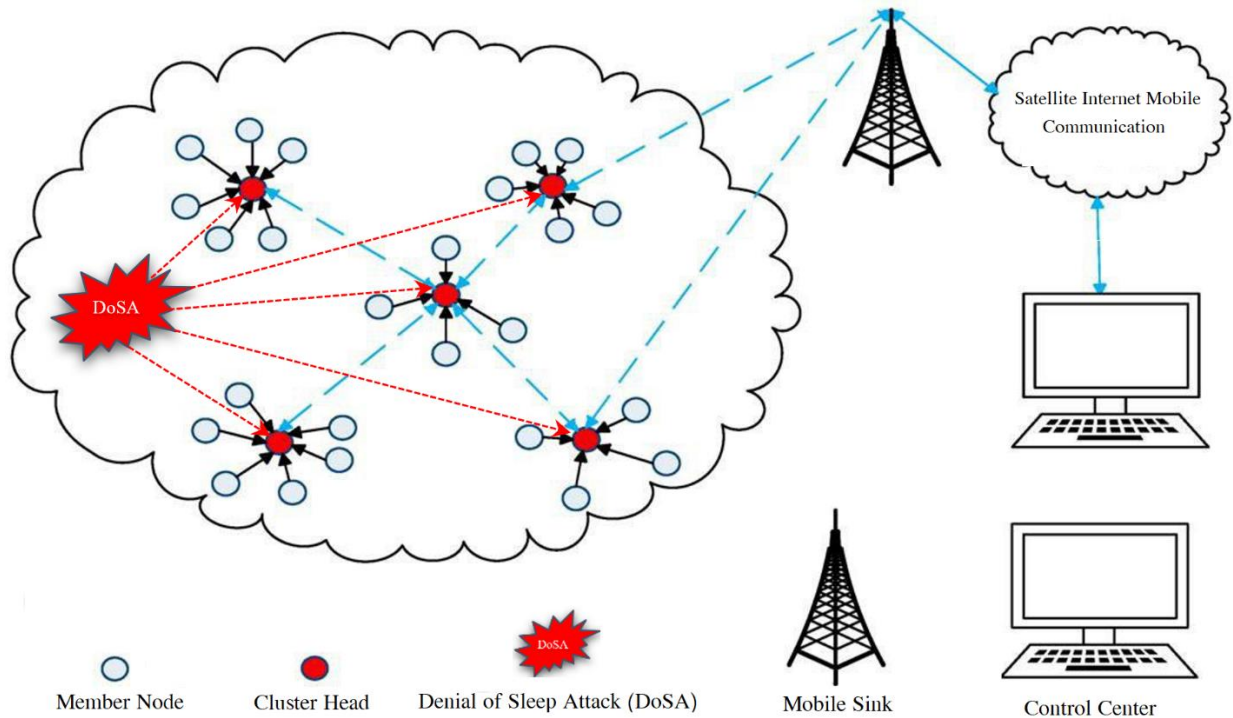

The presented work is structured as the following. Section 2, deals with an overview of basic ideas. In Section 3 the related works are provided on the methods defensive of the system versus denial of sleep attack. Section 4 provides our suggested WSN-FAHN technique details. The parameters used for assessing the performance are studied and simulation outcomes are deliberated in Section 5. Ultimately, the work is concluded in Section 6.

## 2 Overview of basic concepts

Here, some basic issues are briefly provided including the Firefly algorithm and RSA algorithm.

### 2.1 RSA Algorithm

Data aggregation simultaneous and end-to-end confidentiality are provided by employing encryption in terms of privacy homomorphic. Two examples of privacy homomorphism are known as multiplicative *PH* and additive *PH* [8]. Additional processes on encoded information are provided by an additive homomorphic encryption algorithm with no requiring decrypting any separate data. Indeed, the operation costs for increasable *PH* is less than multiplicative *PH*, therefore, using in wireless sensor networks is more appropriate [8].

In addition, multiplicative procedures are supported by a multiplicative homomorphic over encoded data with no decrypting the singular information. $ie. T(a*b) = T(a)*T(b)$. There are two classes of privacy homomorphic supporting cryptographic algorithms such as Asymmetric PH/public key encryption and symmetric *PHs* [8]. In the following section, the RSA cryptography is described:

### 2.1.1 key production of RSA

**Algorithm 1:** Pseudo-code for RSA key production

1: **Initialization:** Let $m$ denote the "modulus"
2: **Initialization:** Let $en$ denote the "encryption exponent"
3: **Initialization:** Let $de$ denote "decryption exponent"
4: **Procedure** key production
5: **Step 1**: $prim1$ and $prim2$, denote two large random of an approximately equal size such that their product $m = prim1 * prim2$ is of the required bit length.
6: **Step 2:** Compute $m = (prim1 * prim2)$ and $\Phi = (prim1-1)(prim2-1)$
7: **Step 3:** Select an integer number $en, 1 < en < \Phi$, thus $\gcd(en, \Phi) = 1$
8: **Step 4:** Compute the secret exponent $de, 1 < de < \Phi$, thus $ed \equiv 1 \bmod \Phi$
9: **Step 5:** $(m, en)$ are public key and $(de, prim1, prim2)$ are private key. Hold all the values $(de, prim1, prim2)$ and $\Phi$ secrets.
10: **End Procedure**

### 2.1.2 Encryption

The pseudo-code for encryption key is shown in Algorithm (2).

**Algorithm 2:** Pseudocode for RSA encryption key

1: **Initialization:** Sender A does the following:
2: **Procedure** Encryption
3: **Step1**: Create the recipient $B's$ public key $(m, en)$.
4: **Step2**: Demonstrate the plaintext message as a positive integer m with $1 < m < en$
5: **Step3:** Calculates the ciphertext $F = m1^{en} \bmod en$
6: **Step4:** Forwards the ciphertext $F\ to\ R$
7: **End Procedure**

### 2.1.3 Decryption

The pseudo-code shown in Algorithm (3) is the decryption key.

**Algorithm 3:** Pseudo-code for RSA decryption key

1: **Initialization:** Receiver B does the following:
2: **Procedure** Decryption
3: **Step1**: $(m, de)$ private key is used to compute $m1 = c^d \bmod m$
4: **Step2**: $m1$ is representative the plaintext from the message
13: **End Procedure**

### 2.2 Firefly Algorithm

Firefly Algorithm (FA) is a multi-modal optimizing algorithm related to the nature-stimulated field. It is inspired by the performance of lightning bugs or fireflies [9]. FA was initially provided by Xin-She at Cambridge University, 2007. FA is experimentally demonstrated to challenge the difficulties more naturally and possesses the potential to over-perform other metaheuristic algorithms. FA is based on 3 basic rules. According to the first, all fireflies are connected to each other with disrespect to gender. In the second rule, it is emphasized that attractiveness is related to the light emission or brightness so that less bright are attracted by bright flies, and the movement is random for the nonexistence of brighter flies. The last main rule indicates that the light emission of the fly is determined or affected by the objective function landscape so that the objective function is proportionate to the brightness [9].

---
**Algorithm 4:** Firefly optimization pseudo-code.

---
Target function $f(x)$

$x_i = (x_1, x_2, x_3, ..., x_d)T$

Produce primary population $x_i = (i = 1, 2, ..., n)$

Let $I_i$ denote light intensity at $x_i$ is specified by $f(x_i)$

Let $\gamma$ denote light absorption coefficient

**while** ($t <$ Max Generation) **do**
  **For** $i = 1:n$ all $n$ fireflies **do**
    **For** $j = 1:i$ all $n$ fireflies **do**
      **If** ($I_j > I_i$) **then**
        Move firefly $i$ towards $j$ in d-dimension;
      **End If**
      Attractiveness varies with distance $r$ via $\exp[\gamma r]$
      Evaluate new approaches and update light intensity
    **End For**
  **End For**
  Grade the fireflies and find the current best
**End while**
Post-process results and visualization

---

The attractiveness among the flies in FA includes two main matters including attractiveness modeling and the different light strengths. Brightness I is formulated as $I(X) \alpha f(x)$ for a particular firefly at location X. However, attractiveness $\beta$ is proportionate to the flies and is associated with the distance $R_{i,j}$ between fireflies $i$ and $j$. Eq. (1) represents the reverse square of light strength $I(r)$ where $I_0$ shows the intensity of the light at the source.

$$I(r) = I_0 e^{-\gamma r^2} \tag{1}$$

Supposing an environment absorption coefficient $\gamma$, the intensity is provided in Eq. (2) where $I_0$ represents the original intensity.

$$I(r) = \left(\frac{I_0}{1 + \gamma r^2}\right) \tag{2}$$

In general, the Euclidean distance is exemplified in Eq. (3), representing the distance between a firefly at the position $X_i$ and another one at a position $X_j$ where $X_{i,k}$ implies the $k^{th}$ component of the spatial coordinate $X_i$.

$$R_{ij} = \|x_i - x_j\| = \sqrt{\sum_{k=1}^{d}(x_{i,k} - x_{j,k})^2} \tag{3}$$

A firefly $I$ was attracted to a brighter one $j$ as shown in Eq. (4), in which the attraction is provided by $\beta_{e^{\gamma r_{ij}^2}}(x_j - x_i)$ and $\alpha\left(rand - \frac{1}{2}\right)$ implies the randomness based on the randomizing parameter $\alpha$.

$$x_i = x_i + \beta_{e^{\gamma r_{ij}^2}} \left( x_j - x_i \right) + \alpha \left( rand - \frac{1}{2} \right) \qquad (4)$$

Additionally, differences in attractiveness are defined by γ affecting the convergence speed and behavior of FA [9].

## 3  Related works

Several works were developed employing various safety measurements to state DOSA and safe WSNs from DoSA. It is not a new problem and it was formerly investigated widely. Different methods were provided by investigators to cope with such attacks.

This study runs the impact of service attack denial resulting in the sleep attack denial in wireless sensor networks utilizing support vector machine learning. A model is established and examined to obtain better results using several kernel functions (sigmoidal, redial, and linear functions). Final results were concluded followed by the severe implementation. Based on the obtained consequences and validating the model, it is concluded that the outstanding throughput is performed by the established system to detect sleep strike attack denial in WSN [10].

A mathematical model is established in [11], to distinguish DoSA attack in WSNs, in terms of absorption of Markov chain, in which, the sensor nodes were found utilizing the Markov chain model. Utilizing this model, attacks are recognized by taking into account the death anticipated time in sensor networks with a usual setup.

In [12], for detecting DoS attack in WSNs, a distributed cooperation-oriented hierarchical framework is used. Within this method, an effective heterogeneous WSN is provided with reliable behavior by finding anomalies in two phases, for minimizing the probability of inappropriate intrusions. For decreasing the attack risk, the networks are isolated in this work from malicious nodes to reject getting fake packets.

In [13], researchers review the denial of service attacks affecting resource accessibility in WSN and their countermeasure by providing a taxonomy.

In heterogeneous wireless sensor networks, a compact hierarchical model specified in [14] is utilized to find sleepless nodes influenced by the attack. In this technique, a cluster on the basis of operative energy is utilized for creating a five-layer hierarchical network to increment its longevity and scalability. In this study, the sensor networks are sectioned into clusters, it is possible to divide them into distinct sections as well. In this method, energy efficiency was attained by the minimisation of the number of active sensors. Furthermore, the sensor nodes are protected by the designed dynamic model from unexpected death. An intrusion detection system is used in this model via power analysis accountable for finding the tasks. Anomaly detecting technique is used in the methods for avoiding intrusion discovery.

In this study, the security threats and vulnerabilities are examined that are forced by the WSNs' distinctive open nature. It summarizes the necessities in WSNs including both the security and survivability issues. Then, a complete survey of different routing and middleware challenges is provided for wireless networks. Then, the potential security threats are detected in the paper at various protocol layers. Here, different security attacks are recognized along with their countermeasures assessed by various investigators recently [15].

The swarm-based defensive method provided in [16] deals with sleep attacks denial using the anomaly detecting model for the determination of the traffic effect within sensor nodes. Therefore, the frequency oscillation technique was established. To gather communication and oscillation frequency, ant factors are utilized as swarm data. The faulty channel is found based on the oscillation frequency; and by isolating the node, the data are obtained, and the damaged channel is eliminated.

In [17], there are two key contributions. First, inspired by the numerous advantages of CSL over ContikiMAC, we trace the present denial-of-sleep defensives for ContikiMAC to CSL. Second, in our work, various security improvement is suggested to these present denial-of-sleep defenses. Operationally, in our work, denial-of-sleep defenses for CSL alleviate well the denial-of-sleep attacks considerably, they also protect against numerous denial-of-sleep attacks compared to the present denial-of-sleep defenses for ContikiMAC. We indicate the accuracy of our denial-of-sleep defenses for CSL analytically and empirically utilizing a whole novel CSL implementing.

In [18], the malicious node is detected in the intrusion detecting system in an isolation table in terms of the attack performance. Abnormal node is found by observing the uncommon performance. The sensor node performance is compared versus the attack performance to cope with irregular data. For the abnormal node performance, it is registered and separated in the isolating table. In Isolation Table Intrusion Detection System (ITIDS), sensor nodes are in charge of observing each other and finding sleep attack denial.

To decrease data flooding and prevent DoS attacks, the storm control mechanism explained in [19] is used. The system traces the received packets frequency, and by exceeding the projected frequency configurating, the node changes the base position and shuts down its wireless receiver for a pre-determined time course.

A two-tier system is offered in [20] in terms of a secure transmission method. In the suggested outline, dynamic session keys are generated using a hash-chain for symmetric encrypting key and doing mutual authentication. This structure needs only some fast and simple calculations in dynamic session key for hash functions, like SHA-1 or MD5. Considering that this outline is unified with the MAC protocol, there are no extra packets compared to the current MAC plans. The appropriate counter measurements are indicated the security analysis versus forge attack and replay attack. Moreover, efficient energy use is demonstrated by analyzing the energy. Furthermore, a potentially novel decision rule is revealed by the detailed energy distribution analysis to cooperate with the security scheme trade-off and energy conservation.

WSN is explained by the National Institute of Standards and Technology (NIST) as "a network consisting of the numerous sensors known as wireless sensors organized in the environment and geographical area". A system consisting of sensors and actuators is known as a sensor network such as some of the calculating elements. WSN includes the sensors nodes sensing the surroundings' physical amounts. Recently, a wireless sensor network has become popular for both civil jobs and the military. Security is one of the most difficult tasks for WSN that is unprotected in an open situation. The cryptographic method is not very proper for protecting the sensor networks from outside attacks [21].

In [22], two enhancements were proposed to Contiki MAC. Dozing enhancement decreases the energy use considerably under ding-dong ditching. In addition, the dozing enhancement aids over normal operation since it decreases the energy use of true wake-ups as well. On the other hand, the secure phase-lock enhancement is a type of Contiki MAC's phase-lock optimizing resisting pulse-delay attacks. Moreover, the secure phase-lock optimizing makes ContikiMAC strong against collision attacks, and further energy effective.

In [23], denial of sleep attack detection (GA-DoSLD) algorithm was suggested by Gunasekaran et al. in terms of an efficient Genetic Algorithm (GA) outline for analyzing the abnormal behavior of the nodes. A Modified-RSA (MRSA) algorithm is run in their algorithm in the base station (BS) for generating and distributing key pairs among the sensor nodes. Before conveying the packets, an optimum route is defined based on the AODV protocol in the sensor nodes. After the determination of the optimum route, the relay node's trustworthiness is guaranteed to utilize fitness computing.

Keerthana and Padmavathi [24] proposed an (EPSO Enhanced Particle Swarm Optimization) method for finding the sinkhole attacks in WSN. In comparison with the present PSO and ACO algorithms, the optimal message drop, packet delivery ratio, false alarm rate, and average delay were provided by the proposed algorithm.

In [25], an algorithm known as Sleep attack Detection Algorithm (SLDA) is suggested for detecting and preventing the attack's denial in the WSNs. This proposed SLDA algorithm is accurate and dynamic for detection of the Sleep attack using trust value, Mobile agent, random key pre-distribution & random password creation. We use a password created randomly and trust value for distinguishing and then for confirming an attacker node and a normal node. Furthermore, this algorithm aids in transmitting the data in a further secured way while preventing the Denial of sleep attacks as well as reducing the power use.

In [26], a two-phase system is provided for detecting wormhole attacks and protecting the DAWA MANET. The suggested procedure includes two stages. Firstly, the efficient routes are chosen in the system through fuzzy logic; then, using an artificial immune system, the immune route is determined in the chosen routes. In this work, making the wormhole tunnels via out-of-band high-power channels leads to the identification of the efficient wormhole attacks. Though, a prerequisite deliberation exists in this work indicating that when the wormhole attacks influence the destination nodes, no packets are received that is essentially not true.

An IDS (Intrusion Detection System) was provided in [27], to protect against the security problems through the human immune system (HIS). The IDSs are utilized for detecting and responding to efforts to cooperate with the target system. As the UASs act in the real world, the validation and testing these systems with various sensors are opposed to the problems. This design is stimulated by HIS. In mapping, insecure signals are equal to an antigen detected by antibody-based training designs and eliminated from the operation cycle. The fast detecting of the intrusive signals and quarantining their activity are amongst the key usages of the proposed design.

## 4 The proposed WSN-FAHN approach

In this section, we propose a DoSA-immune schema by applying the Firefly, Hopfield Neural Networks (HNN), and RSA optimization. The WSN-FAHN comprises of five sections, such as the Assumption of the WSN-FAHN Method is discussed in Sect, 4.1. Cluster formation is considered in Sect, 4.2. In section 4.2 describes the sending data to CHs. Section 4.4 describes determine the movement of sink points. Data authentication by the mobile sink is discussed in Sect. 4.5.

The Flowchart of the WSN-FAHN is shown in Fig. 2.

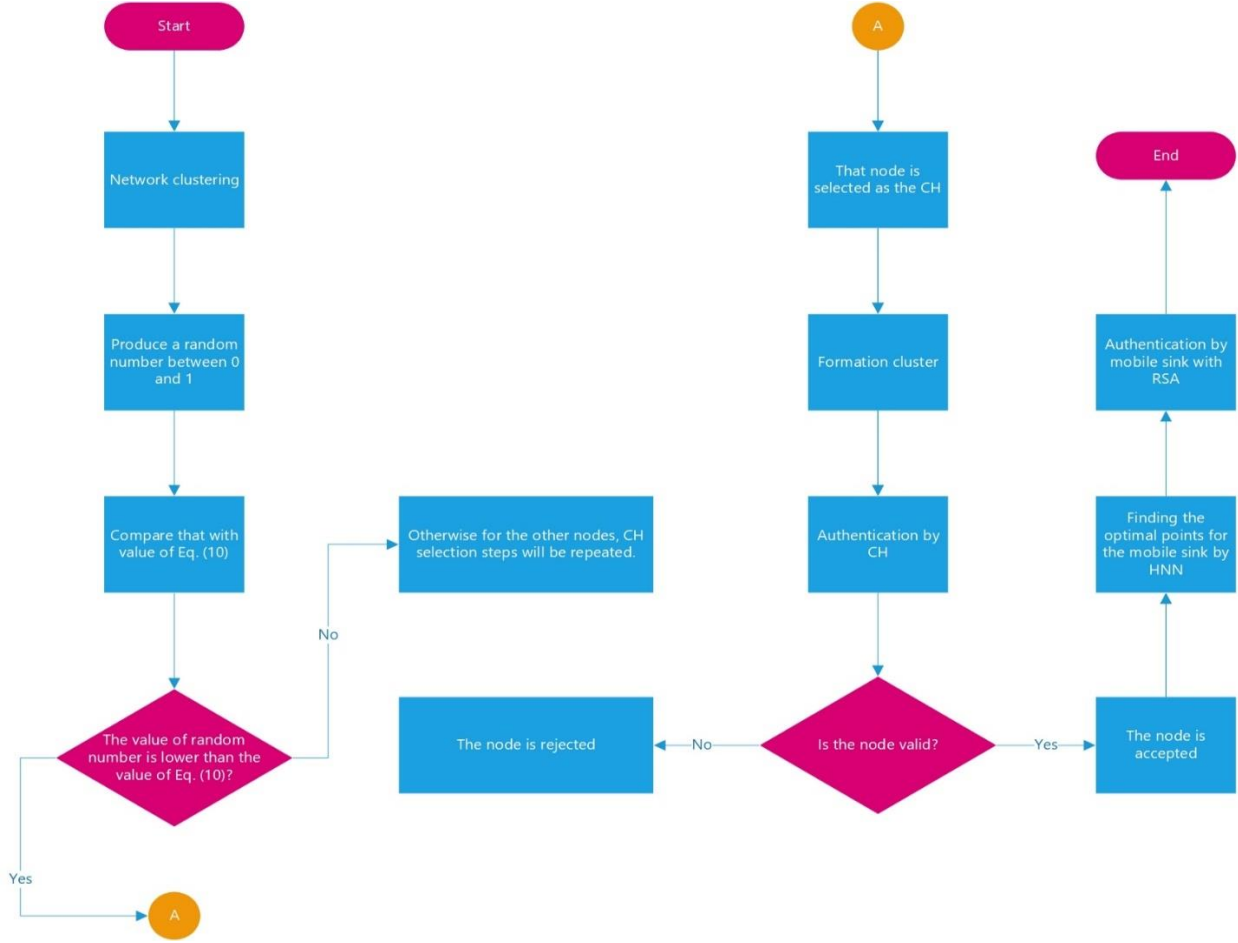

**Fig. 2** WSN-FAHN flowchart.

### 4.1 Assumption of the WSN-FAHN Method

The WSN in the proposed WSN-FAHN consists of nodes distributed randomly. We bring up the following characteristics for our WSN-FAHN. The network has a multi-channel mobile sink and n sensor nodes which have primary energy $E_o$. The communication of cluster members with CH is done in one hop. Moreover, each CH communicates with the sink in one hop. Nodes are able to adjust their transmission radius. Also, the network activity cycle is divided into several rounds. $T_{round}$ is the time spends in each rounds and it is calculated by Eq. (5).

$$T_{round} = \left(T_{cf} + T_{dc} + T_{dr}\right) \tag{5}$$

$T_{cf}$ is the time needed to form cluster. $T_{dc}$ is the time of sending data of member cluster to CH node. $T_{dr}$ is the time needed to send data of CH node to the mobile sink. The proposed method has four phases as: cluster formation phase, sending data to CH, Determine the movement of sink points, and data authentication by the mobile sink.

### 4.2 Cluster formation

The sink aggregates the percentage of required clusters and information of all sensor nodes for the entire broadcast network. After receiving information from the sink, all nodes produce a random number between zero to one and compare it to $T(n)$ value that is calculated according to Eq. (6) [28].

$$T(n) = \begin{cases} \dfrac{z}{1-z\left(q \bmod \left(\dfrac{1}{z}\right)\right)} & N1 \in G \\ 0 & otherwise \end{cases} \quad (6)$$

If the produced number is less than its value, that node is chosen as the CH node in the current step. The selected headers compute their intensity by the target function according to Eq. (7) [28] and send it to other common nodes as a message form. Then, all CH nodes keep the maximum intensity value which are calculated by common nodes for current step.

$$I(x) = \left(\dfrac{I_o}{\left(1 + \gamma S_i^2\right)}\right) \quad (7)$$

The smaller the distance among two nodes, the greater the intensity quantity. It means that intensity value has an inverse relationship with its distance. $I_o$ is the initial value of intensity in each node. $S_i$ value is computed by Eq. (8).

$$S_i = \left(S_i + \beta\left(S_j - S_i\right) + \alpha\left(round - 0.5\right)\right) \quad (8)$$

In Eq. (8), $S_i$ is the location of CH sensor and $S_j$ is the location of common sensors. Also, common nodes calculate their intensity value by Eq. (7). Also, they keep the maximum intensity values that are related to CH nodes. Then, common nodes compere their intensity values to received intensity value from CH nodes and connect to CH node which has the maximum intensity value.

**4.3 Sending data to cluster heads**
In this phase, information is gathered from the environment by common sensor nodes and they are sent to CH node.
**Authentication at cluster level:** The sleep cycle of the sensors is arranged by using S-MAC protocol in sensor networks. S-MAC protocols work by sending synchronize signals to set the sleep cycle of the nodes. These protocols work with using control messages such as Request Forwarding (RTS) and Ready to Sending (CTS) that are known as synchronization packets. Repeating control packets as RTS message is one of the effective ways to prevent sleep deprivation attack. This causes that nodes do not fall to sleep. Therefore, their power loses. When these messages are sent over a short period of time, network nodes will not have an enough opportunity to go to sleep mode and come back again. This leads to lose battery power. This lose can occur for all nodes at attacker 's transmission time. Attackers send synchronization messages that means how much time a node can stay at sleep mode. Our proposed method is used authentication at two levels. One level is applied by CHs to examine member nodes of the clusters and other level is applied by the mobile sink. Also, in the first step, we define a threshold parameter for synchronization packets. During the normal operation, if synchronization value is lower than threshold value, synchronization is not used authentication. When the synchronization is higher than threshold quantity, sleep deprivation attack is likely happen. Thus, synchronization should be applied for authentication. In addition, synchronization packets are authenticated to synchronize when S-MAC algorithm is running. Synchronization packet must insert its ID in the first field before time sleep begin. When CH nodes receive a packet, they investigate whether the sender of the packet exists in its cluster. If the sender does not exist in its cluster, synchronization packet is not accepted, hence this is the first level of authentication. If the first level of authentication is successful, CH node which receives synchronization packet calculates arrival time from the node. So, each CH node maintains calculated time in its storage system. If this arrival time is more than the arrival time threshold, it may occur sleep deprivation attack. In this situation, we cannot block the suspicious node because the attack may be occurred by an external attacker that knows the node behavior. Its aim is to disturb network operation by sending packets with different IDs. In our WSN-FAHN, there is an authentication mode that is activated by CH nodes that detects suspicious behavior as passing the desired threshold. Each CH node sends a sync-authentication packet to all other nodes in its cluster to change their mode to authentication mode and forces nodes to send the authentication token to the next sync token.

After authentication of node data at the cluster level, each CH creates a TDMA scheduler and sends it to all cluster members which have been authenticated. When TDMA is fixed in all clusters, all the common nodes within each cluster send their data to their specific TDMA, and each CH integrates them after receiving all the data from its members then send to the mobile sink. Each CH is able to store data in its buffer until the sink stops in suitable area near its neighbor.

**4.4 Determine the movement of sink points**

In this phase, Hopefield Network finds some optimal points for mobile sink in the network. Hopefield Network has several neurons that their activation shows optimal points for mobile sink. In HNN, the number of neurons belong to $T_{dr}$ value. $T_{dr}$ is the time of CH data to the mobile sink. Figure 3 shows the combination of HNN and WSN.

**Fig. 3** Combination of HNN and WSN.

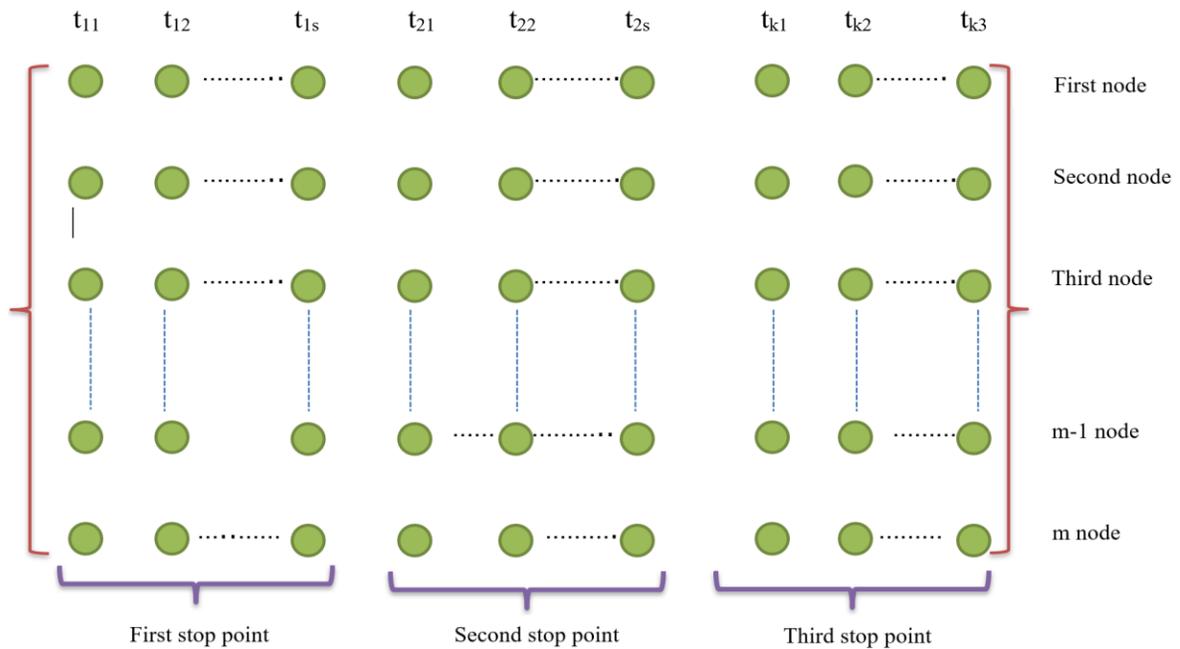

The number of HNN rows equals to number of CH nodes in the sensor network (M) and the number of HNN columns equals to K*S (K is the number of stop points) and S is the time unit number that the sink can stop at any points. Also, T is the sink allowed time in any columns.

**Consider $T_{dr}$ time in HNN:** First, $T_{dr}$ time should be determined. In our proposed method, all CH nodes are able to store data in their buffers until the sink stops in a suitable area near them. In this method, the sink has the maximum time to gather data from CH nodes. This time is shown by $T_{dr}^{max}$ which is computed with Eq. (9) and Eq. (10) [28].

$$T_{dr}^{max} = \left( \varphi * \sum_{i=1}^{M} \left( \frac{C_i}{r_i} \right) \right) \tag{9}$$

$$\varphi = \left( \xi T_{dc} d_s \right) \tag{10}$$

$C_i$ is the number of cluster member $i'th$. $d_s$ is the rate of received data from the environment by the cluster members. $\xi$ is the coefficient of data integration. $r_i$ is the data transfer rate based on bits per second in the CH $i'th$ and $M$ is the number of CH nodes. Note that it is important that sink stop time should not exceed more than $T_{dr}$ quantity. Now, we get the number of active neurons for one row or a CH node. When a neuron is active in $X$ row and $Y$ column, it means that $X$ CH node is sending data to the sink at the $Y$ stop point. To send all data of a CH node, certain number of neurons is needed that is computed by Eq. (11) [28].

$$neron_{on} = \frac{(\varphi * C_i)}{(T * r_i)} \tag{11}$$

After obtaining the number of active neurons in a CH, the total number of neurons (active and inactive) should be calculated in the entire HNN. We calculate the total number of neurons for $T_{dr}^{max}$.

**Calculate all neurons for $T_{dr}^{max}$:** Considering the maximum time for $T_{dr}$ means that the sink can have one or more stop points for gathering data from each cluster nodes. Therefore, it multiplies number of active $neron$ ($neron_{on}$) of CH to number of stop points of CH ($neron_{on} * k$). Also, to get all network $neron$, equation multiplies to $M$ which is number of CH. This time is shown by $neron(T_{dr}^{max})$ which is computed with Eq. (12).

$$neron(T_{dr}^{max}) = neron_{on} * k * m \tag{12}$$

### 4.5 Data authentication by the mobile sink
CH sends data which are buffered for mobile sink after authentication at cluster level and receiving data from CH node. The CH sends data to mobile sink as soon as the sink stops its neighborhood. Then, mobile sink authenticates all data sent by CH and receives data from CH. Therefore, the overall process is after clustering, choosing proper CHs, and authentication at cluster level, mobile sink receives all data from sensor nodes as CHs and all other nodes. Then, it stores them to access them.

The transfer of the keys from mobile sink to other nodes is a vital task because keys may be attacked. In order to protect keys from attacks, data is transferred by interlocking protocol that keys are encrypted by AES algorithm. First part is transmitted. After receiving a reply from received node, the second part is transmitted. If these two parts are connected each other, the key can decode in the receiving BS. In order to perform this transmission, nodes which are in the networks must agree with the key symmetric encryption method. Also, encrypted key is divided into two parts at inter locking protocol. We have used AES algorithm that divides cryptography into two parts. These two parts are performed respectively. First, the node is authenticated. If it is valid and the node is the member of the cluster, the second part of the key is sent. The public and private keys created with RSA algorithm are transferred by interlocking protocol. The nodes which send synchronization sleep messages are proving nodes and receiving nodes acts as evaluator. Each node has a unique private key that we name it F. Both proving node and evaluator node shares public key. Moreover, the mobile sink sends the proving encrypted key from the BS when the evaluator requests it. The mobile sink calculates $H=F^2 \bmod V$ instead of sending key directly. In the equation, F is private key and V is public key. H value is sent to evaluator whenever it requests it. Receiving mobile sink decodes key when it receives a reply from node and if data of the CH node is valid, it is accepted by the mobile sink.

The Pseudo-code for our proposed WSN-FAHN is shown in the algorithm (5).

| | |
|---|---|
| | **Algorithm (5):** Pseudo code for WSN-FAHN proposed schema |
| 1: | **Step 1. Network Clustering** |
| 2: | Select the cluster head (CH); |
| 3 | Produce a random number between 0 and 1 by nodes; |
| 4: | Compare the random number with value of **Eq. (10)**; |
| 5: | **If** the value of random number is lower than value of T(n) **Then** |
| 6: | Its sensor selects to CH; |
| 7: | **Else** |
| 8: | **Repeat** the steps 3 to 5 to other nodes to selecting CH; |
| 9: | **Step 2. Formation cluster** |
| 10: | Determining the value of intensity function for selecting CH by **Eq. (11)**; |
| 11: | Sending that to normal nodes; |
| 12: | Determining the value of intensity function for normal nodes by **Eq. (12)**; |
| 13: | Calculating **Eq. (12)** to getting the location of CH nodes and normal nodes; |
| 14: | Compare the value of intensity function of CH nodes with the value of intensity function for normal nodes; |
| 15: | Normal nodes connecting to clusters head with high value of intensity function; |
| 16: | Sending a connection request to CH; |
| 17: | Formation cluster; |
| 18: | **Step 3. Sending data to cluster node** |
| 19: | Node Authentication at cluster level; |
| 20: | Calculate Threshold; |
| 21: | Calculate Interval; |
| 22: | **While** Interval > Threshold **do** |
| 23: | **For** All of the node at cluster level |
| 24: | The node is invalid (attacker node); |
| 25: | **For** All of the node at cluster level |
| 26: | Should be done Authentication Phase; |
| 27: | **End For** |
| 28: | **End For** |
| 29: | Node is Accepted |
| 30: | **End while** |
| 31: | **Step 4. Finding the optimal points for the mobile sink;** |
| 32: | Finding maximum point to stop sink for each CH to collect data from each CH by **Eq. (13) & Eq. (14)**; |
| 33: | Calculate the number of active neurons for each CH to finding the optimal points to moving the sink from **Eq. (16)**; |
| 34: | Sending data from CH to mobile sink by giving results from **Lines 32, 33**; |
| 35: | **Step 5. Authentication by mobile sink;** |
| 36: | RSA key production by mobile sink; |
| 37: | Key distribution using by the interlock protocol; |
| 38: | Node authentication is done using by node reply; |
| 39: | **If** the node is invalid **Then** |
| 40: | The node is rejected. |
| 41: | **Else** |
| 42: | The node is Accepted |
| 43: | **End If** |

# 5 Performance evaluation

In following section, we show and discuss the experimental simulation results of the WSN-FAHN to prevent DoSA.

### 5.1 Performance metrics

In following section, we investigate the performance and effectiveness of the WSN-FAHN by a numerical simulation in NS-2. The results are compared with CrossLayer and GA-DoSLD approaches proposed in [20] and [23], respectively. The notations of the considered for WSN-FAHN are as Table 1.

**Table 1** Notations description

| Parameters | Description |
|---|---|
| T | Network throughput |
| PDR | Packet Delivery Rate |
| DR | Detection rate |
| RE | Let RE denote the "Residual energy" |
| NL | Let RE denote the "Network lifetime" |
| $X_i$ | Let $X_i$ denote the "received packets by node $i$" |
| $Y_i$ | Let $Y_i$ denote the "sent packets by node $i$" |
| $P_S$ | Let $P_S$ denote the "Packet size" |
| $S_P$ | Let $S_p$ denote the "Simulation stop time" |
| $S_T$ | Let $S_T$ denote the "Simulation start time" |

### 5.1.1 T

Eq. (13) show the *T* for *n* experiments , and is calculated in Kbps [29]. This metric is obtained via dividing the amount of packet received in the destination by the packet arrival time.

$$T = \left(\frac{1}{n}\right) * \left(\frac{\sum_{i=1}^{n} X_i * P_s}{S_p - S_T}\right) * \left(\frac{8}{1000}\right) \tag{13}$$

### 5.1.2 PDR

The average ratio of the number of successfully received data packets to the number of data packets sent. The PDR obtained for *n* experiments is shown by Eq. (14) [30-34].

$$PDR = \left(\frac{1}{n}\right) * \left(\frac{\sum_{i=1}^{n} X_i}{\sum_{i=1}^{n} Y_i} * 100\%\right) \tag{14}$$

### 5.1.3 NL

Prolong the lifetime of the network is the main goal of WSNs. And this by reducing the energy consumption of the sensor nodes is possible According to the explanation, the *NL* is the passed time between of communication and sensing commencement with the receiver, and the time in which the final communication link from active sensor to the receiver is broken. Eq. (15) denote the calculation of the WSN lifetime.

$$NL = \sum_{i=1}^{m} LCH_i \qquad \text{Where} \qquad LCH_i \text{ is the lifetime of } i\text{th } CH. \tag{15}$$

### 5.1.4 RE

When transmitter sensor nodes transmit information and when receiving sensor nodes receive information, energy is consumed. The total remaining energy in a sensor node can be measured via Eq. (16) based on the transmission and reception times. The parameters used for the *RE* are listed in Table 2.

**Table 2** *RE* parameters

| Parameters | Description |
|---|---|
| $di_0$ | Let $di_0$ denote "reference distance greater than the Fraunhofer-distance" |
| $di$ | Let $di$ denote "the distance on which the packet is sent" |
| $Lb$ | Let $Lb$ denote "the number of bits per packet" |
| $di^2$ | Let $di^2$ denote "power loss of free space channel model" |
| $di^4$ | Let $di^4$ denote "power loss of multi-path fading channel model" |
| $E_{elec}$ | Let $E_{elec}$ denote "amount of energy getting dissipated during sent or receive" |
| $lb \in fsi$ | Let $lb \in fsi$ denote the "transmission efficiency" |
| $lb \in mpi$ | Let $lb \in mpi$ denote the "condition of the channel" |

$$Energy_{residual} = Energy_{initial} - \{ET_X + ER_X + E_{sys}\} \quad \text{Where} \tag{16}$$

$$ET_X(1b, di) = \{lbE_{elec} + lb\varepsilon_{fs}di^2, di < di_0\} \tag{17}$$
$$= \{lbE_{elec} + lb\varepsilon_{mpi}di^4, di \geq di_0\}$$

Energy consumed [35] are calculated using Eq. (17), and Eq. (18), respectively.

$$ER_X = lbE_{elec}. \tag{18}$$

The Energy consumed parameter is define as: $\begin{cases} E_{elec} = 90nJ/bit, \\ \varepsilon_{fsi} = 30 pJ/bit/m^2, \\ \varepsilon_{mpi} = 0.0023 pJ/bit/m^4 \end{cases}$

### 5.1.5 DR

It is determined as the ratio of the number of DoSA sensors marked to the total number of existing DoSA sensors in the WSN. *DR* is calculated by Eq. (19). Table 3 show the parameters for *DR*.

**Table 3** *DR* parameters

| Parameters | Description |
|---|---|
| True Positive (TP) | The *TP* is calculated by the total number of marked the DoSA sensors divided by the total number of the DoSA sensors. |
| False Positive (FP) | The *FP* is calculated by the total number of sensors wrongly detected as the DoSA sensors divided by the total number of normal sensors. |
| True Negative (TN) | The rate of the DoSA sensor that were correctly marked as a DoSA sensor. |
| False Negative (FN) | The rate of the DoSA sensor to total normal sensors that were mistakenly marked as a normal sensor. |

$$DR = \left(\frac{TP}{TP+FN}\right)*100 \quad \text{where} \tag{19}$$

$$TPR = \left(\frac{TP}{TP+FN}\right)*100 \quad ; \quad FN = \left(\frac{TP+TN}{All}\right)*100 \quad All = TP+TN+FP+FN$$

$$TN = \left(\frac{TN}{TN+FP}\right)*100 \quad ; \quad FP = \left(\frac{FP}{FP+TN}\right)*100$$

## 5.2 Simulation Setup and Comparing Algorithms

The basic advantage of simulation is shortening analysis, largely in large-scale networks [36, 37]. All simulations were performed using Network Simulator2 (NS-2) tools.

## 5.3 Simulation results

We have simulated WSN-FAHN schema in the NS-2 on Linux Ubuntu 18.04 LTS. The parameters used in WSN-FAHN simulation is given in Table 4.

**Table 4** Parameters used in WSN-FAHN simulation.

| Parameters | Value |
|---|---|
| Topology | $90 \times 90\ m^2$ |
| Sensor nodes | 300 |
| Range of transmission | $150 - 250\ m$ |
| Size of packets | $512\ Bytes$ |
| $E_{elec}$ | $100 nJ / bit$ |
| $\varepsilon_{fsi}$ | $20 pJ / bit / m^2$ |
| $\varepsilon_{mpi}$ | $.0015 pJ / bit / m^4$ |
| Energy | $45\ J$ |
| Idle power | $51\ mW$ |
| Receiving power | $55\ mW$ |
| Sleep power | $35\ \mu W$ |
| Power of transmission | $51\ mW$ |
| Sensing power of sensors | $8*10^{-8}\ J$ |
| Simulation time | $70\ s$ |

The performance of WSN-FAHN with that of CrossLayer and GA-DoSLD in terms of $DR$, $T$, $PDR$, $Energy_{residual}$, $NL$ are shown in Table 5-9.

**Table 5** $DR$ vs misbehaving sensor ratio.

| Misbehaving sensor ratio | $DR$ (%) | | |
|---|---|---|---|
| | $CrossLayer$ | $GA-DoSLD$ | $ASDA$ |
| 0 | 100 | 100 | 100 |
| 0.05 | 85 | 95 | 98 |
| 0.15 | 80 | 91 | 97 |
| 0.25 | 75 | 85 | 96 |
| 0.35 | 72 | 81 | 95 |

**Table 6** $T$ vs misbehaving sensor ratio.

| Misbehaving sensor ratio | $T$ (in kbps) | | |
|---|---|---|---|
| | CrossLayer | GA − DoSLD | ASDA |
| 0 | 900.85 | 900.85 | 900.85 |
| 0.05 | 750.8 | 700.4 | 870.5 |
| 0.15 | 580.7 | 610.9 | 835.4 |
| 0.25 | 550.7 | 580.4 | 800.2 |
| 0.35 | 420.03 | 490.3 | 780.9 |

**Table 7** $PDR$ vs misbehaving sensor ratio.

| Misbehaving sensor ratio | $PDR$ (in %) | | |
|---|---|---|---|
| | CrossLayer | GA − DoSLD | ASDA |
| 0 | 100 | 100 | 100 |
| 0.05 | 82 | 92 | 94 |
| 0.15 | 71 | 85 | 91 |
| 0.25 | 65 | 81 | 89 |
| 0.35 | 60 | 78 | 88 |

**Table 8** $Energy_{residual}$ vs misbehaving sensor ratio.

| Misbehaving sensor ratio | $Energy_{residual}$ (in %) | | |
|---|---|---|---|
| | CrossLayer | GA − DoSLD | ASDA |
| 0 | 100 | 100 | 100 |
| 0.05 | 75.8 | 81.7 | 95.7 |
| 0.15 | 71.8 | 78.4 | 91.4 |
| 0.25 | 67.8 | 74.2 | 87.4 |
| 0.35 | 65.8 | 71.08 | 82.8 |

**Table 9** $NL$ vs misbehaving sensor ratio.

| Misbehaving sensor ratio | $NL$ (in sec) | | |
|---|---|---|---|
| | CrossLayer | GA − DoSLD | ASDA |
| 0 | 1000 | 1000 | 1000 |
| 0.05 | 690 | 878 | 954 |
| 0.15 | 610 | 790 | 904 |
| 0.25 | 500 | 710 | 897 |
| 0.35 | 450 | 650 | 887 |

The average values of all methods under DoSA is given in Table 10.

**Table 10** Average values.

| Schemes | | Throughput | PDR | Detection rate | Residual energy | Lifetime |
|---|---|---|---|---|---|---|
| GA – DoSLD | Number of sensor nodes | 542.416 | 76.4 | 81.55 | 81 | 617.2 |
| | Misbehaving sensor ratio. | 640.616 | 75.6 | 82.4 | 76.24 | 650 |
| | Simulation times | 417.8571429 | 77.71428571 | 76.71428571 | 80.3 | 28.41666667 |
| | Attack interval | 284.0525 | 66.75 | 70.875 | 72.125 | 618.75 |
| CrossLayer | Number of sensor nodes | 612.588 | 89.4 | 85.928 | 85.774 | 650 |
| | Misbehaving sensor ratio. | 656.57 | 87.2 | 90.4 | 81.076 | 805.6 |
| | Simulation times | 557.2857143 | 81.85714286 | 84.57142857 | 84.81428571 | 30.01428571 |
| | Attack interval | 364.9825 | 82.875 | 76.5 | 76.1225 | 642.625 |
| ASDA | Number of sensor nodes | 722.686 | 94.4 | 92.09 | 93.08 | 875 |
| | Misbehaving sensor ratio. | 837.57 | 92.4 | 97.2 | 91.46 | 928.4 |
| | Simulation times | 827.14286 | 90.285714 | 92.142857 | 93.185714 | 37.23 |
| | Attack interval | 602.75 | 89 | 94.25 | 86.785 | 854.375 |

Figure 4 presents the average throughput of sensors participating in data sending/receiving operations in WSN-FAHN, GA-DosLD, and CrossLayer methods based on network topology. The results obtained from simulations indicate that in the WSN-FAHN method, since the CH sends its data to the sink only when the sink stops at an appropriate location in its vicinity, it leads to decreased energy consumption in the sensors and reduced data transmission time. Also, throughput of the WSN-FAHN increases compared to GA-DosLD and CrossLayer methods.

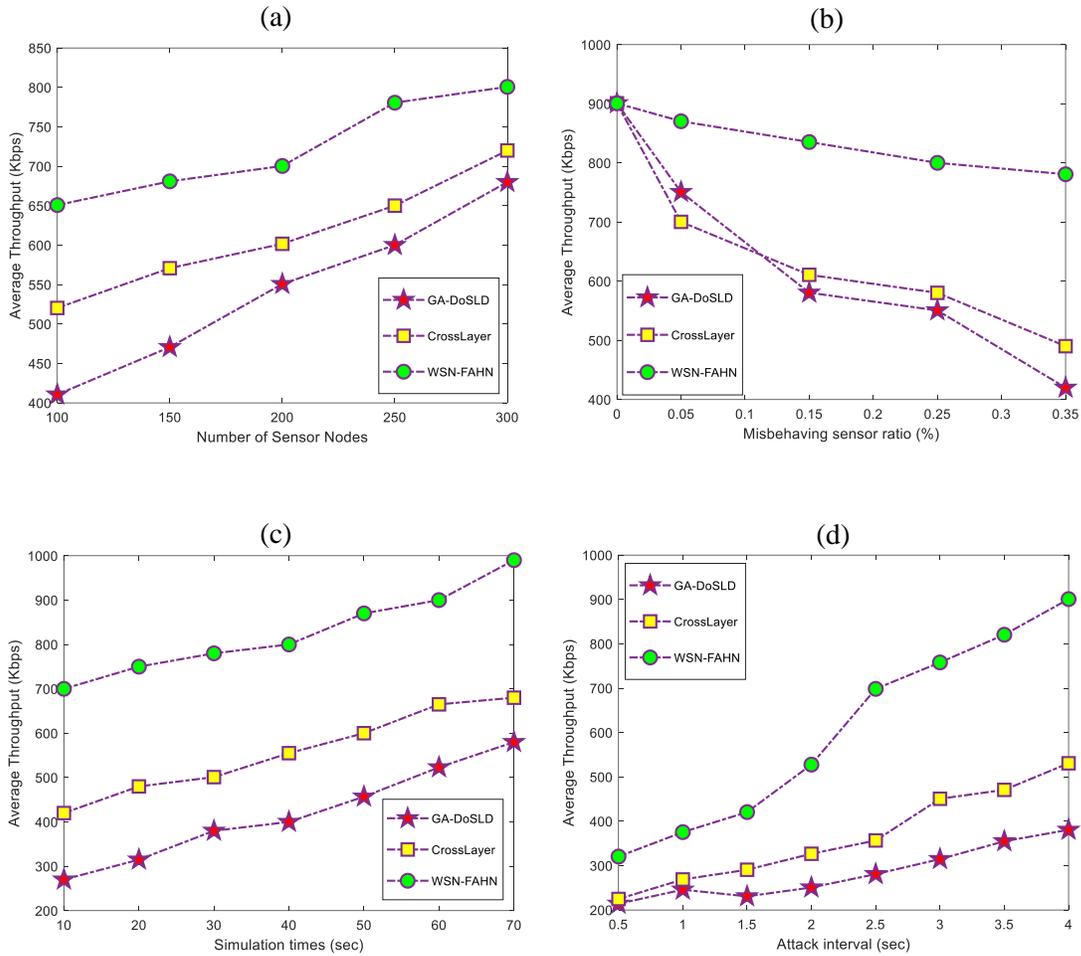

**Fig. 4** Comparison of the WSN-FAHN, CrossLayer and GA-DoSLD models in term of Throughput.

Figure 5 presents the PDR of sensors participating in the sending/receiving operations in the proposed and other methods. Results obtained from simulation show that the packet reception rate in the proposed WSN-FAHN method is higher because, due to mobility of the sink, cluster head data reach the sink in the least amount of time possible. This leads to lower lost data and transmission of data time to the sink and higher data reception rates in the proposed method compare to GA-DosLD and CrossLayer methods. Also, to supply some description on these results, it is worth mentioning that when the power is properly adjusted, our proposed scheme requires less power to send packets. Plus, in each CH iteration selection, the node residual energy and its distance to the receiver criterion is considered, and the node with the highest level of energy is chosen as the CH. This will control the energy available in the network and increase the remaining energy, resulting in a better performance.

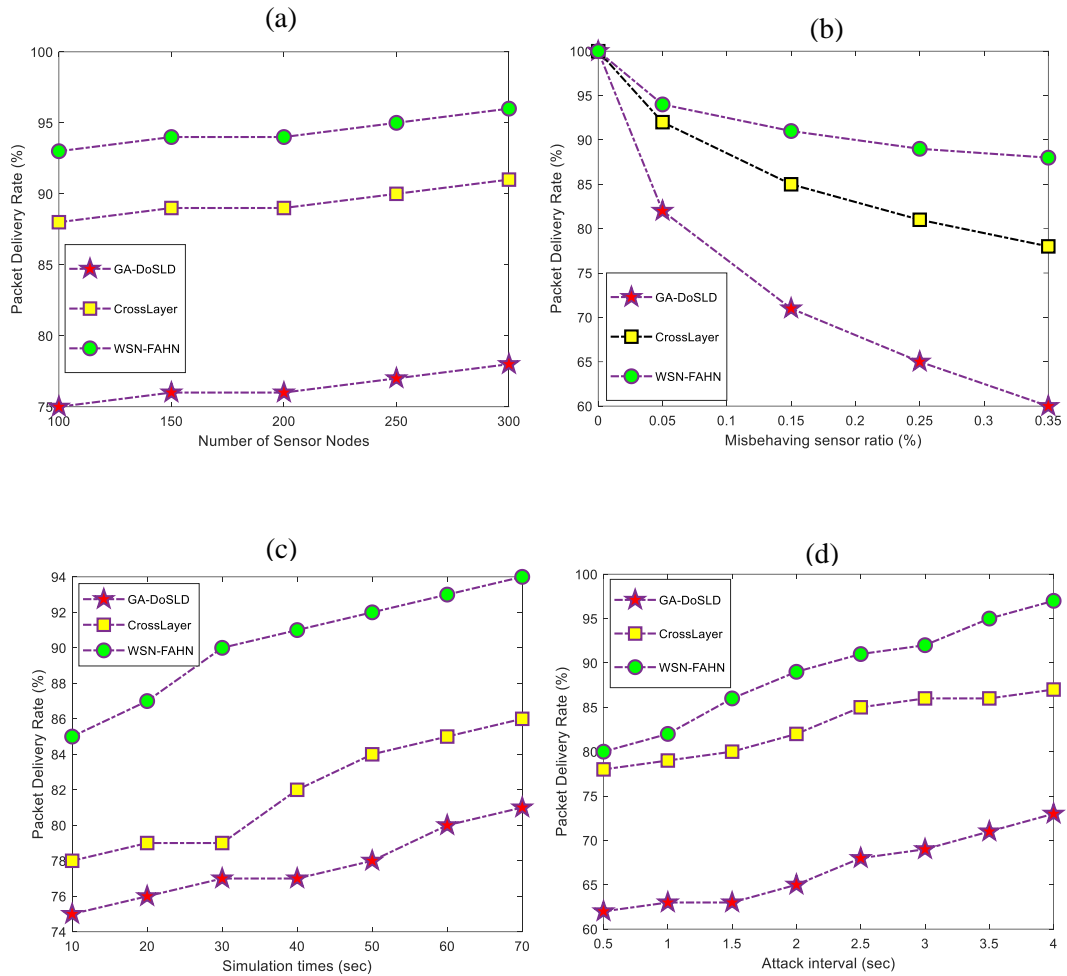

**Fig. 5** Comparison of the WSN-FAHN, CrossLayer and GA-DoSLD models in term of PDR.

Comparing DR in the proposed WSN-FAHN method with GA-DosLD and CrosLayer methods, according to Figure 6, shows that the detection rate is higher in the proposed method compared to GA-DosLD and CrossLayer methods. Since in the proposed method, an authentication step is carried out by CH for all sensors first and then, due to the decreased distance of sending data to the sink by the CH, packets reach the sink faster and node energies do not decrease. This makes the sink carry out the second authentication faster and with higher detection rate.

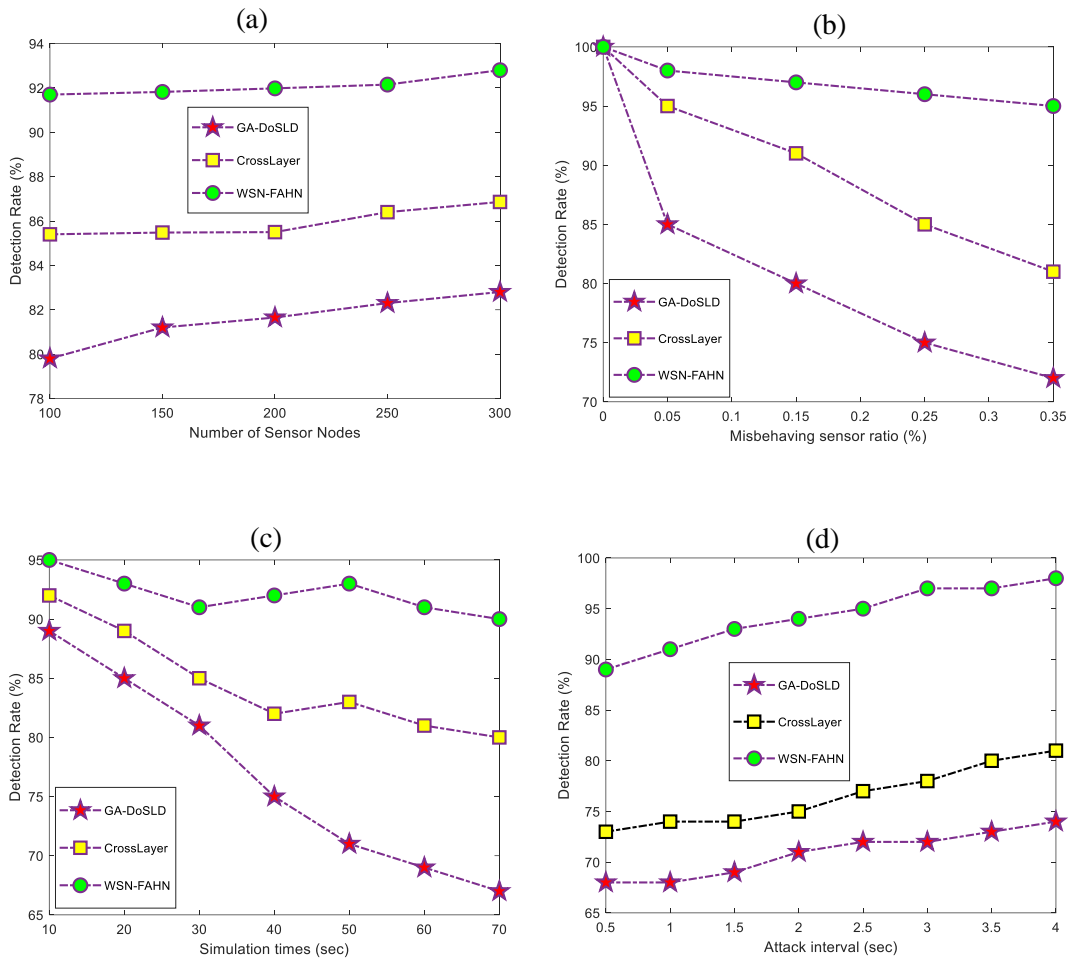

**Fig. 6** Comparison of the WSN-FAHN, CrossLayer and GA-DoSLD models in term of Detection rate.

Figure 7 presents the residual energy values of WSN-FAHN and other methods for different values. The results show that residual energy in the WSN-FAHN method does not deplete rapidly due to the mobility of sink and low data transmission distance. This leads to higher residual energy compared to other methods like GA-DosLD and CrossLayer where sink is stationary and long distances must be passed to transmit cluster head data to sink. As demonstrated in the Fig. 6, WSN-FAHN decreases the *RE* by more than 20 and 12% those of GA-DoSLD and CrossLayer, respectively.

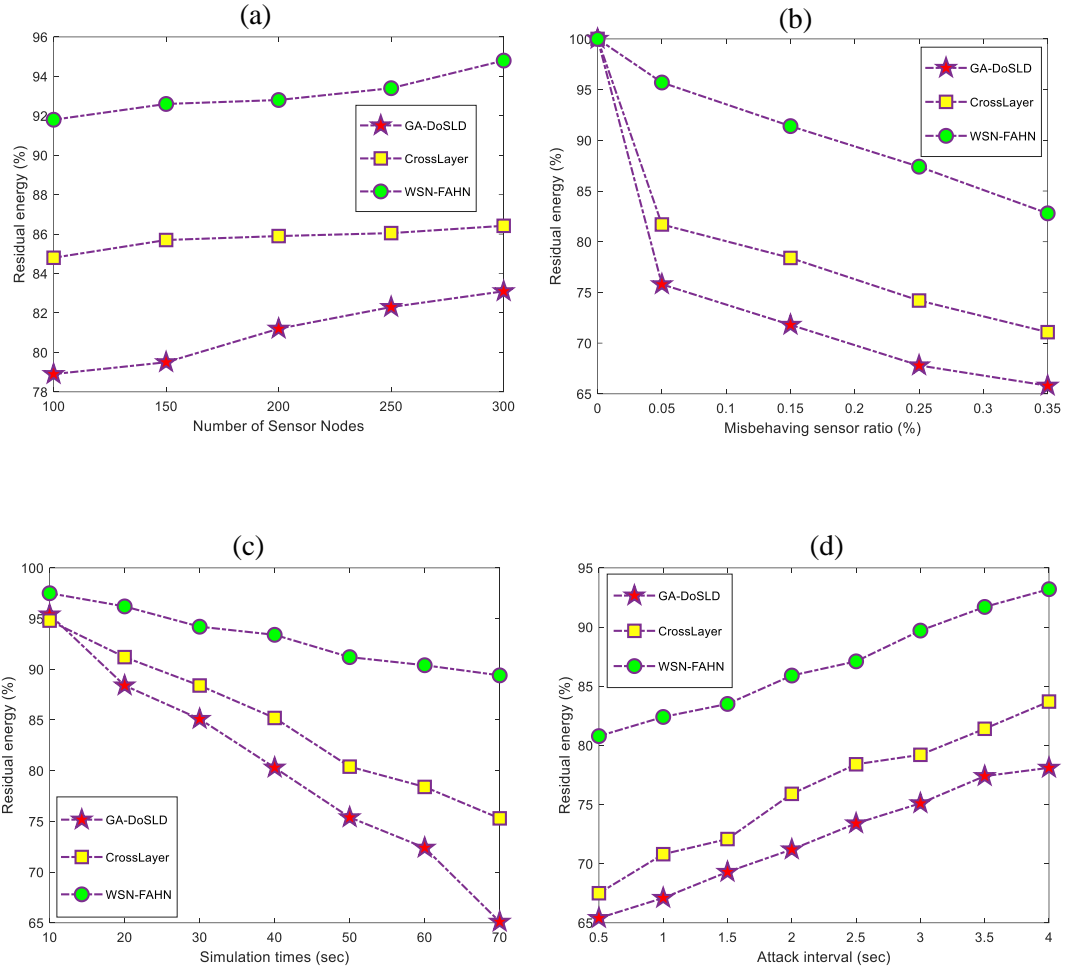

**Fig. 7** Comparison of the WSN-FAHN, CrossLayer and GA-DoSLD models in term of Residual energy.

Figure 8 presents the lifetime comparison of the proposed method with existing GA-DosLD and CrossLayer clustering techniques. The results extracted from simulations indicate that sink mobility in the proposed WSN-FAHN method leads to reduced data transmission distance from cluster head to sink and therefore lower energy consumption in the sensors. This makes network lifetime longer compared to GA-DosLD and CrossLayer methods which have stationary sinks.

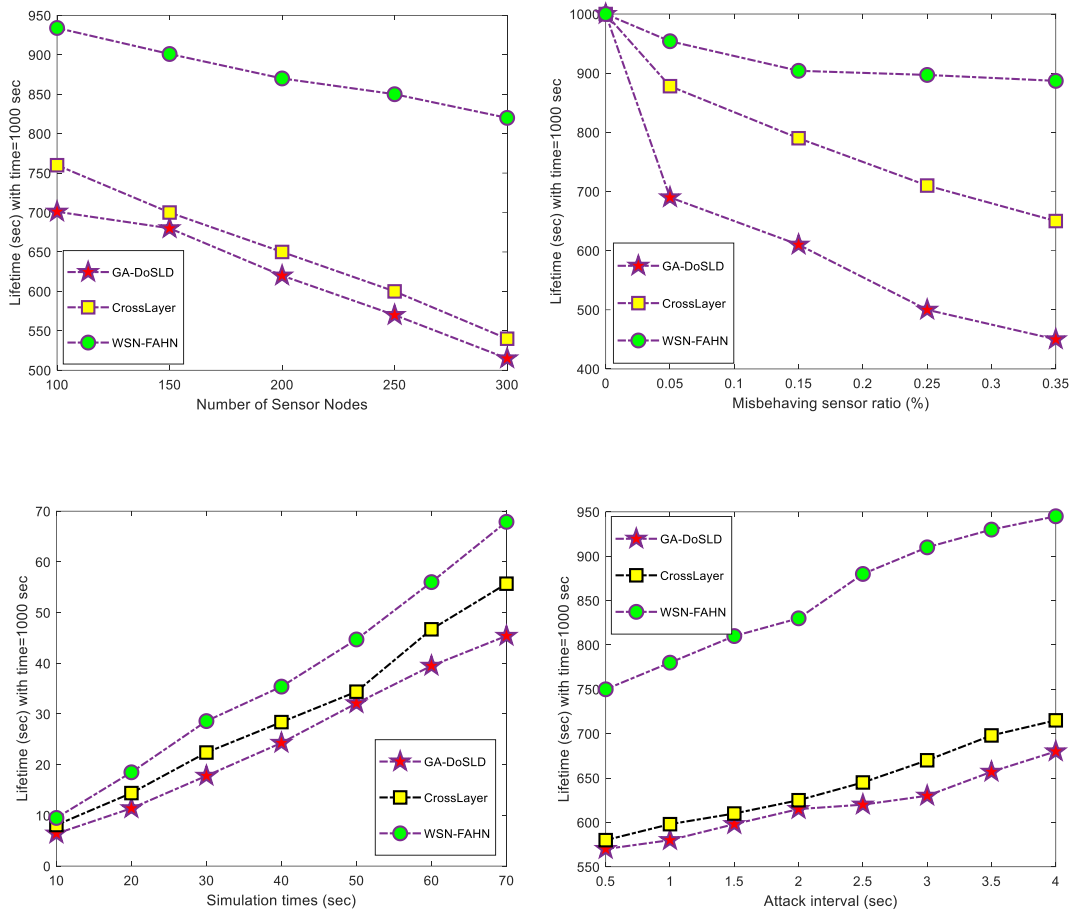

**Fig. 8** Comparison of the WSN-FAHN, CrossLayer and GA-DoSLD models in term of Lifetime.

Figure 9, show the comparison between WSN-FAHN, RSA (128), WSN-FAHN, RSA (256), WSN-FAHN, RSA (512), WSN-FAHN, RSA (768), WSN-FAHN, RSA (1024), WSN-FAHN, RSA (1280) cryptography algorithms. The encryption and decryption times increase in all methods with the increase in key size and chunk size. By increasing key size, we get better security in all methods but speed decreases in the meanwhile. In comparison, WSN-FAHN with key size 512, encryption time is 254 ms and decryption time is 789 ms with block size 1024. Also, for WSN-FAHN with key size 1024, encryption time is 324 ms while decryption time is 1280 ms with block size 1024. Therefore, WSN-FAHN performs well compared to other encryption and decryption methods in terms of encryption and decryption. Also, WSN-FAHN has significantly improved security. In addition, in the RSA algorithm, the longer key length is, more time will be spent on key encryption and decryption by the nodes. This will in turn lead to more energy consumption in the nodes.

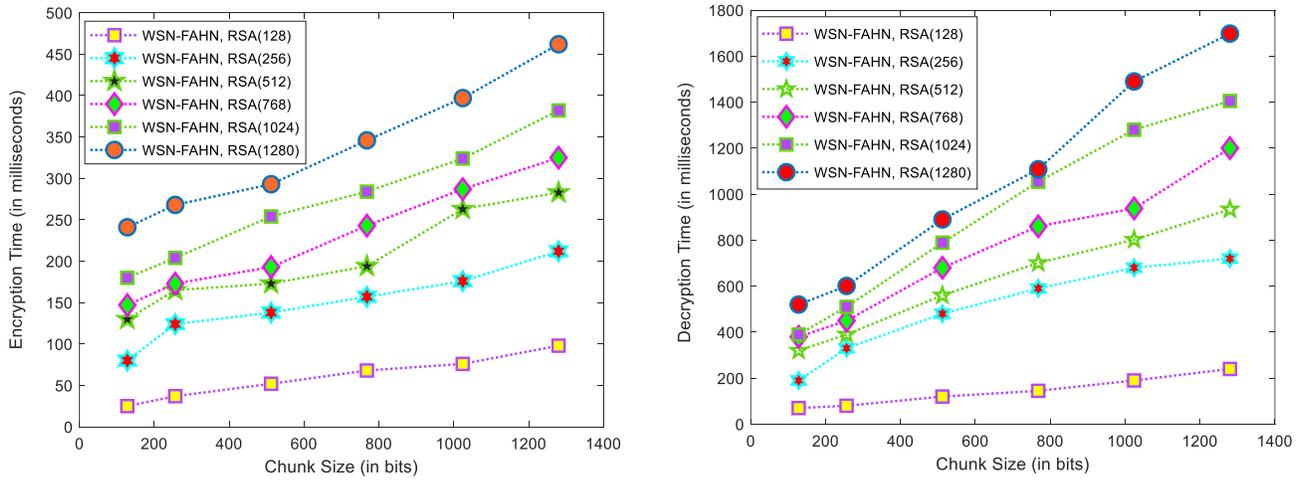

**Fig. 9** Encryption and Decryption time vs different chunk size in WSN-FAHN

The comparison results of the WSN-FAHN, in terms of consumption energy at different type of data size and different chunk Size are provided in Figure 10.

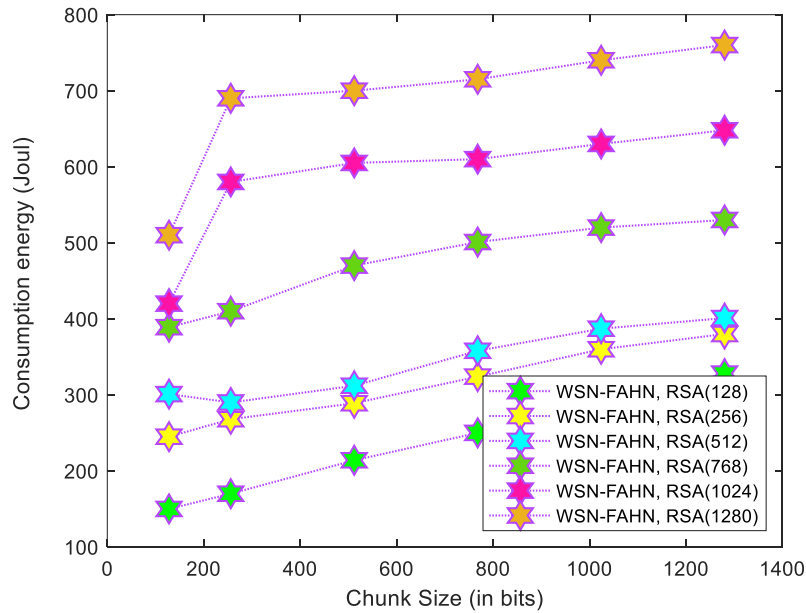

**Fig. 10** Consumption energy vs different chunk size in WSN-FAHN

## 6 Conclusion and Future work

In this paper, a hybrid approach is proposed based on mobile sink, firefly algorithm based on leach, and HNN. Thus, mobile sink is applied to both improve energy consumption and increase network lifetime. Firefly algorithm is proposed to cluster nodes and authenticate in two levels to prevent from DoSA. In addition, HNN detects the direction route of the sink movement to send data of CH. The proposed solution effectively prevents DoSL attacks, since all the nodes that transmit synchronisation messages should be validated, if not, before the messages are accepted. If the nodes are not validated, the messages will be rejected. The attacker node is unable to replay the sleep synchronisation signal, since the sleep schedule cannot be accepted without authentication, and sensor nodes possess limited resources and capabilities. Employing the authentication method proposed here extends the network lifetime, through effective battery consumption while maintaining a secure communication within the network. The results confirmed that our WSN-FAHN scheme is capable of exhibiting high-levels of security and high ratio of detection (exceeding 93.92%). It addition, our proposed scheme has high PDR (more than 91.52%), high throughput (more than 90.75%), high Average Residual energy (more than 91.125%), and high network lifetime (more than 89.95%), in comparison with the other approaches currently being employed. In future work, the use of several mobile sinks is suggested to further reduce energy consumption in wireless sensor networks.